\documentstyle[12pt]{article}
\newcommand{\beq}{\begin{equation}}
\newcommand{\eeq}{\end{equation}}
\newcommand{\bi}{\bibitem}
\newcommand{\om}{\omega}
\begin{document}
{\hbox to\hsize{August, 1997 \hfill TAC-1997-028}
\bigskip
\vglue .06in
\begin{center}
\vglue .06in
\begin{center}
{\Large \bf {Velocity of signal in attractive potential \\
and propagation of light in gravitational field
  }}
\end{center}
\bigskip
{\bf A.D. Dolgov}
 \\[.05in]
{\it{Teoretisk Astrofysik Center\\
 Juliane Maries Vej 30, DK-2100, Copenhagen, Denmark
\footnote{Also: ITEP, Bol. Cheremushkinskaya 25, Moscow 113259, Russia.}
}}\\
\bigskip
{\bf I.B. Khriplovich }
 \\[.05in]
{\it{ Budker Institute of Nuclear Physics\\
  630090 Novosibirsk, Russia
 }}\\[.40in]

\end{center}
\begin{abstract}

The propagation of a massless field in attractive and repulsive potentials
is considered.  It is shown that though the group velocity in such potentials
can be larger than one, the wave front propagates with the speed of light.
A larger-than-one group velocity leads only to a strong deformation of the wave
packet. The results obtained are applied to the light propagation in a
gravitational field. An erroneous assertion concerning the last problem,
recently made in the literature, is refuted.

\end{abstract}
\bigskip

It is well known that the theory of a free field with a negative mass squared,
$m^2 < 0$, is unstable with respect to small perturbations and, moreover,
has no ground state. Though it is sometimes argued that a signal in such
theories would propagate with a super-luminal velocity, it does not make much
sense. In models of spontaneous symmetry breaking the instability introduced
by the mass term $m^2 \phi^2$ (with $m^2 <0$) in the potential is "cured" by
nonlinear interaction terms, for instance by $\lambda \phi^4$. It gives rise
to a stable vacuum state and to a positive mass over this vacuum.

It is interesting to consider what would happen to a massless particle
propagating through an attractive potential. For a well localized wave packet,
with the size which is small in comparison with the characteristic potential
range $a$, such a potential is equivalent to a negative mass squared in a
finite space region. We study here the problems of stability and of the speed
of signal propagation in such a potential. These questions are not of
pure methodological interest but refer in particular to the physical process
of light propagation in an external gravitational background.
The Maxwell equations in the empty space can be written as [1-3]:
\beq
F^{\;\;\;\;\;\;\; ;\nu}_{\mu\lambda;\nu} +R_{\mu\lambda\rho\nu}F^{\rho\nu} = 0
\label{max}
\eeq
After diagonalization we obtain from eq.~(\ref{max})
separate equations for the polarization eigenmodes, which describe the
propagation of a massless field in gravitational background.

We will model our problem by the Klein-Gordon equation:
\beq
-\partial_t^2 \phi = -\Delta \phi + \mu^2 (x) \phi
\label{kg}
\eeq
Positive $\mu^2 $ describes a repulsive potential, while $\mu^2 <0$ describes
an attractive one. We consider for simplicity the one-dimensional problem,
which in three dimensions corresponds to the forward scattering.

The group velocity in the potential depends upon its sign
and is smaller than unity in a repulsive potential for $\omega^2 > \mu^2$
and larger than unity in
an attractive potential. This can be easily seen in the  eikonal
approximation (which is valid if the signal frequency $\om$ is much larger
than both $|\mu|$ and the inverse effective range of the potential, $1/a$
($\mu^2 / \om^2 \ll 1$, $\om a \gg 1$).
A wave packet moving in such a potential can be described as
\beq
\phi (x,t) = \int_{-\infty}^\infty d\om B(\om -\om_0)
\exp \left( -i\om t + i\om x -{i\over 2 \om} \int_{-\infty}^x dx' \mu^2 (x')
\right)
\label{eik}
\eeq
The function $B(\om -\om_0)$ is by assumption strongly peaked near the
central frequency of the packet $\om_0$. Expanding the exponential near
$\om = \om_0$, we get
\beq
\phi (x,t) =
e^{  -i\om_0 t + i\om_0 x -
i \int_{-\infty}^x dx' \mu^2 (x') /2\om_0 }
A \left[ t-x -{1\over 2\om_0^2} \int_{-\infty}^x dx' \mu^2 (x') \right]
\label{eik1}
\eeq
The function $A$ describes the shape of the wave packet. Its maximum moves
with the velocity
\beq
v_g =  1 - \mu^2 (x) / (2\om_0^2).
\label{vg}
\eeq
Thus for
$\mu^2 > 0$ the group velocity, $v_g$, is smaller than the speed of light,
while for $\mu^2 < 0$ it is larger than the speed of light.
We will show however that the velocity of the front propagation of the signal
remains equal to one for both repulsive and attractive potentials. It means
in particular that the wave packet is strongly distorted when propagating
in "attractive" media, so that its maximum reaches its front. After that
the velocity of the propagation of the maximum is equal to that of the
front. The shape of the packet becomes very sharp and the usual expressions
for the group velocity, in particular (\ref{vg}),
which refer to sufficiently smooth packets becomes
non-applicable. A similar phenomenon takes place in optics in the region of
anomalous dispersion (see e.g. \cite{anom}).

For a proper approach to the problem of the velocity of signal propagation one
has to start with an initial signal which has a sharp front. In other
words, the signal should be strictly absent at initial time $t=0$ at distances
larger than some $x=x_0$. For simplicity we take $x_0 = 0$ so that the wave
function does not contain the non-essential phase $\exp (i\omega x_0)$. The
region where the potential is essentially nonzero, is assumed to be far
to the right from $x=0$. We will expand the solution of eq.~(\ref{kg}) in
terms of its eigenmodes $\phi_\omega (x)$:
\beq
\phi (x,t) = \int^\infty_{-\infty} d\omega C(\omega ) \phi_\omega (x)
e^{i\omega t}
\label{phixt}
\eeq
We will take $\phi_\omega (x)$ in the form which describes an incoming
and reflected waves at large negative $x$ and transmitted wave at large
positive $x$. Accordingly it has the following asymptotic behavior:
\begin{eqnarray}
\phi_\omega (x)= e^{i\omega x} + d(\om)e^{-i\omega x}\;\;\;\; ({\rm for} \;\;
x\rightarrow -\infty ),\nonumber \\
\phi_\omega (x)=  f(\om)e^{i\omega x} \;\; ({\rm for} \;\;
x\rightarrow \infty  ).
\label{phiom}
\end{eqnarray}
To satisfy the above mentioned condition of absence of the incoming wave at
$x>0$ for $t<0$, the coefficient function $C(\om)$ should be analytic in
the upper-half plane of complex $\om$. Indeed, in this case the contour of
integration in $\om$ can be closed in upper half-plane for $(x-t) > 0$ and
the corresponding integral vanishes.

In three-dimensional case we are interested in the forward scattering
amplitude $f(\om)$. Its analytical properties follow directly  from
the analysis of the operator
\beq
H= -\Delta + U ( r) \equiv -\Delta + \mu^2 (r),
\label{h}
\eeq
as is discussed in textbooks on quantum mechanics (see e.g. \cite{ll,ar}).
The amplitude is analytic in the upper half-plane $\om$ and for sufficiently
strong attractive potentials has poles on the positive imaginary axis which
correspond to bound states. In one-dimensional case
the analytical properties of the functions
$d(\om)$ and $f(\om)$ are the same (see e.g. \cite{four,msm1}),
with the only difference that an attractive one-dimensional potential
has usually at least one bound state and
respectively at least one pole on the upper imaginary axis.

Let us start from the study of the wave front propagation for the repulsive
potential. The amplitude of the transmitted wave is given by eq.~(\ref{phiom})
where the product $C(\om) f(\om)$ is analytic in the upper half plane. Thus
by the same reason that the incoming wave vanishes for $x>t$, the transmitted
wave vanishes in the same interval and the super-luminal propagation is
impossible. We came effectively to the known statement \cite{leon} that the
velocity of the wave front is determined by the asymptotics of the
refraction index $n(\om)$ as $\om \rightarrow \infty$. In turn this
asymptotics is obviously governed by the characteristics of the wave
equation which are not influenced by the potential. Thus the velocity of the
wave front propagation does not exceed the speed of light (in the case
considered it is equal to the speed of light), though the group velocity
generally differs from unity. The group velocity may be larger
than speed of light but the wave front goes with the speed of light.

We are considering here the problem of propagation of signal in attractive
potentials or, if in repulsive ones, then well above the barrier.
As to the case of the
barrier penetration by a signal, strong deformation of the wave
packet results here in nontrivial phenomena. However, their
discussion is beyond the scope of the present article.
A detailed treatment of the problem of sub-barrier propagation of a signal as
well as the list of appropriate references, both theoretical and experimental,
is presented in ref.~\cite{msm2}, where the general expression for the
time-delay of the signal for sub-barrier motion has been calculated.

Let us turn now to the case of attractive potential. When the potential is
sufficiently weak, the scattering amplitude, as mentioned above, is analytic
in the  upper half-plane and the problem of signal propagation does not differ
at all from the already considered case of repulsive potential. In a strong
potential the operator $H= -\Delta + \mu^2 (x)$ acquires negative eigenvalues,
$\om^2_j<0$, which correspond to imaginary $\om_j$ and to unstable eigenmodes,
$\phi_\om \sim \exp (|\om_j | t)$. Small fluctuations in such a potential
exponentially grow up and the initial problem with a signal absent for $x>0$ is
not so trivial from the physical point of view, as it is above in the stable
case. Still we can consider the following mathematically well defined problem,
namely the propagation of the signal with sharp front,
$\phi (x,0) = 0 $ for $x>0$ assuming that the rising fluctuations are absent.
We will present the wave function in the same form as above, eq.~(\ref{phixt}),
but with the integration done along the contour which goes over
the singularities in the upper half-plane (see Fig.~1). Let us emphasize that
this choice of the contour is necessary for the meaningful formulation
of the problem, with the signal absent for  $x>0$. One
can see that again the signal propagates with the speed of light,
i.e. $\phi (x,t) =0$ for $x>t$.

It can be seen that the signal propagates without essential deformation
till it reaches the region of the potential. Here it excites the unstable modes
which also propagate with the speed of light, but exponentially grow behind the
wave front, $ \phi \sim \exp |\om_j (x-t) |$. The amplitudes of these unstable
modes are equal to the coefficients of decomposition of the incoming wave in
terms of the unstable eigenmodes of the operator $H$.
There are of course other rising contributions which are not related to
the propagating wave. They originate from the unstable
fluctuations (noise) in the region of non-vanishing potential.

One can however distinguish experimentally the propagating signal, which may
carry information, from the unstable exponentially rising noise. Let us assume
that the initial signal is a wave packet with a high central frequency
$\om_0 \gg |\om_j|$. To carry some information, this signal should be modulated
with the frequency $\tilde\om $ which is also much larger than $|\om_j|$.
Of course the noise coming from the unstable fluctuations may reach a detector
faster than the signal and at first sight can mimic an acausal propagation.
However the detector can distinguish between them.
For example the detector may be constructed in such a way that it does not
register the slowly varying  background. If it is sensitive only to a
high frequency signal and if the measuring time is small in comparison with
$1/|\om_j |$ and large in comparison with $1/\om_0$ and $1/\tilde \om$,
the detector would react only  to causally propagating signal.

Coming back to the problem of the propagation of electromagnetic waves
in a gravitational background, we see that neither the "potential" term
(proportional to $R_{\mu\nu\rho\lambda}$) nor the terms with the Christoffel
symbols implicit in $F_{\mu\nu;\lambda}^{\;\;\;\;\;\;;\lambda} $, influence the
characteristics of the wave equation and therefore do not change the velocity
of the wave front propagation which remains equal to the speed of light. (It
sounds funny: "speed of light is equal to the speed of light", but the meaning
is clear.) However the radiative corrections to the Maxwell equations
change the second-derivative terms and correspondingly do change
the characteristics of the wave equation \cite{radcor,ibk}. This can make
in particular the velocity of wave front propagation larger than one. The
criticism of this conclusion made in ref. \cite{mp} is based on the erroneous
identification of the signal velocity with the group velocity instead of the
wave front one.

Let us mention in conclusion that strong attractive fields indeed exist
in nature. We mean the gravitational fields of black holes at $r\sim r_g$.
The vacuum instability in these fields results in the Hawking radiation.

\bigskip

{\bf Acknowledgment.}

This work was supported in part by Danmarks Grundforskningsfond through its
funding of the Theoretical Astrophysical Center (TAC).
I.Kh. thanks TAC for hospitality and acknowledges the
support by the Russian Foundation for Basic Research through grant
No. 95-02-04436-a.

\bigskip

{\bf Figure caption.}

Integration contour in the complex $\om$-plane in eq. (\ref{phixt}).

\end{document}